\documentclass{aa}  
\usepackage{graphicx}
\usepackage{txfonts}
\usepackage{multirow}
\usepackage{verbatim}
\usepackage{xcolor}

\newcommand{\gx}{GX~9+1}
\newcommand{\ixpe}{{IXPE}\xspace}
\newcommand{\nus}{{NuSTAR}\xspace}
\newcommand{\integral}{{INTEGRAL}\xspace}
\newcommand{\maxi}{{MAXI}\xspace}
\newcommand{\nicer}{{NICER}\xspace}  
  
\newcommand{\bat}{{Swift/BAT}\xspace}

\xspaceaddexceptions{+ - *}
\begin{document}

   \title{X-ray spectro-polarimetry analysis of the weakly magnetized neutron star X-ray binary GX 9+1 }


   \author{A. Tarana
          \inst{1}
          \and
          F. Capitanio \inst{1}
          \and
          A. Gnarini \inst{2}
          \and
          S. Fabiani \inst{1}
          \and
          F. Ursini \inst{2} 
          \and
          S. Bianchi \inst{2} 
          \and
          C. Ferrigno \inst{3}    
          \and
          M. Parra \inst{4}
          \and
           M. Cocchi \inst{5}
          \and
          R. Farinelli \inst{6}
          \and
           G. Matt \inst{2}
          \and
          P. Soffitta \inst{1}
          \and 
         A. Bobrikova\inst{7}
          \and 
           P. Kaaret\inst{8}
        \and 
        M. Ng\inst{9,10}
        \and 
          J. Poutanen\inst{7}
        \and
        S. Ravi\inst{11}
          }

   \institute{INAF Istituto di Astrofisica e Planetologia Spaziali, Via
del Fosso del Cavaliere 100, 00133 Roma, Italy \\
              \email{antonella.tarana@inaf.it} 
         \and
            Dipartimento di Matematica e Fisica, Università degli Studi Roma
            Tre, via della Vasca Navale 84, 00146 Roma, Italy 
             \and 
             Department of Astronomy, University of Geneva, Ch. d’Ecogia 16,
            CH-1290 Versoix, Geneva, Switzerland 
             \and
            Department of Physics, Ehime University, 2-5 Bunkyocho, Matsuyama, Ehime 790-8577, Japan 
            \and 
             INAF -- Osservatorio Astronomico di Cagliari, Cagliari, Italy  
            \and
             INAF -- Osservatorio di Astrofisica e Scienza dello Spazio, Via P. Gobetti 101, 40129 Bologna, Italy 
            \and
            Department of Physics and Astronomy, FI-20014 University of Turku, Finland 
            \and
            NASA Marshall Space Flight Center, Huntsville, AL 35812, USA 
            \and
            Department of Physics, McGill University, 3600 rue University, Montr\'{e}al, QC H3A 2T8, Canada 
            \and
            Trottier Space Institute, McGill University, 3550 rue University, Montr\'{e}al, QC H3A 2A7, Canada 
            \and
            MIT Kavli Institute for Astrophysics and Space Research, Massachusetts Institute of Technology, Cambridge, MA 02139, USA 
             }

   \date{Received XX XX, 2025; accepted XX, XX, 2025}

 
  \abstract
   {We present an X-ray spectro-polarimetric study of the weakly magnetized neutron star low mass X-ray binary \gx, utilizing data from the Imaging X-ray Polarimetry Explorer (\ixpe), alongside simultaneous \nus, \nicer and \integral observations. \gx, located in the Galactic bulge, is a persistently bright Atoll source known for its spectral variability along the color--color diagram. Our spectral analysis during the soft state confirms emission dominated by a soft blackbody and thermal Comptonization components, with no evidence of a hard X-ray tail. Moreover, these observations suggest a relatively low--inclination system ($23\degr<i<46\degr$) with a weak reflection component, consistent with emission from the accretion disk and neutron star boundary layer. Spectro--polarimetric analysis reveals no significant polarization in the 2--8 keV range, with a 3$\sigma$ level upper limit of the polarization degree of 1.9\%. However, marginal evidence of polarization was detected in the 2--3 keV band at the 95.5$\%$ confidence level (2$\sigma$), suggesting potential contributions from scattering effects in the individual spectral components (disk, reflection and Comptonization) that could cancel each other out due to the different orientations of their polarization angles. This behavior aligns with other Atoll sources observed by \ixpe, which typically exhibit lower and less variable polarization degrees compared to Z--class sources. 
   }

   {}


   \maketitle
%

\section{Introduction}~\label{sec:intro}
Weakly magnetized neutron star (WMNS) are Low-mass X-ray binaries (LMXBs) systems hosting as compact object a low magnetic field NS \citep[see] [for a review]{DiSalvo.etAl.2023, Done.etAl.2007}. They are typically classified into two main categories: Atoll sources and Z sources based on their X-ray spectral and timing properties \citep{Hasinger.VanDerKlis.1989, VanDerKlis.2006}. Atoll and Z sources show a different track in the color-color diagram (CCD), a C-like and Z-like shape respectively, which reflects different spectral changes. 
In particular, Atoll sources exhibit two primary spectral states: the island state (IS) and the banana state (BS), which correspond to distinct accretion regimes and luminosity levels \citep{Hasinger.VanDerKlis.1989}. 
~IS is characterized by lower X-ray luminosities, typically in the range of $L_x\sim10^{36}-10^{37}$ erg s$^{-1}$ and a harder X-ray spectrum dominated by Comptonized emission from a hot electron corona or a boundary layer near the NS surface. The banana state (BS), on the other hand, occurs at higher X-ray luminosities ($L_x\sim10^{37}-10^{38}$ erg s$^{-1}$), is associated with relatively high accretion rates. The emission in this state is softer and dominated by thermal radiation, typically modeled as a blackbody or a multi-temperature disk blackbody component originating from the NS surface or the inner accretion disk \citep{Iaria.2005, Done.etAl.2007}. The banana branch is further divided into the lower banana, where the spectrum is softer, dominated by thermal disk and neutron star emission, and the upper banana, where the spectrum hardens slightly due to a residual Comptonized component \citep{Lin2007}. 
However, the Atoll and Z classification is no longer considered entirely rigid, as some sources have been observed to undergone a transition between Atoll and Z-like behavior, such as XTE J1701–462 \citep{Homan2007, Lin2009} and GX 13+1 \citep{Fridriksson_2015, Schnerr_2003}. These findings suggest that Atoll and Z sources may represent different accretion regimes of the same underlying physical process, rather than two strictly separate categories \citep{Muno2002ZAtoll, Gierlinski2002}. In fact, Z sources persistently emit at high luminosity ($\sim 10^{38}$ erg s$^{-1}$), close to the Eddington limit for a NS, while the Atoll sources are persistent and transient sources that can become rather bright, but typically show luminosities of few tenths of the Eddington limit ($ 10^{36} - 10^{38}$~erg s$^{-1}$).

In this framework, \gx~has historically been classified among the so-called Bright Atoll sources or GX-Atoll sources, along with GX 3+1, GX 9+9, and the peculiar GX 13+1, which are persistently bright and remain almost exclusively in the banana brach \citep[][and references therein]{Iaria.2005, Iaria_2020, VanDerKlis.2006}. Discovered in 1965, this source, located in the Galactic bulge, has been the subject of numerous studies up to the present day. Early X-ray observations from missions such as EXOSAT, BeppoSAX, NUSTAR and AstroSAT characterized the spectral behavior of \gx, confirming it is predominantly observed to move only along the banana branch of the CCD, with a high X-ray luminosity in the range of $\sim 10^{37}-10^{38}$ erg s$^{-1}$ \citep{Thomas2023, Iaria.2005, Langmeier1985}. 
Spectral studies often model \gx~using two components: a soft blackbody component that originates either from the NS’s surface or from the inner accretion disk,  and a hard component associated with thermal Comptonization from seed photons being scattered by a hot corona or spreading layer \citep[see for example][]{Iaria.2005}. \gx~does not exhibit a significant hard X-ray tail extending to high energies, which is often observed in other NS LMXBs \citep[see][]{Paizis2005}.

The near infrared (NIR) counter part of \gx~was identified thanks to precise localization provided by Chandra. It is consistent with the secondary being a late-type dwarf, likely an M-type star, and is probably located in front of the Galactic bulge at a relatively low distance, about 4 kpc \citep{vanderBerg2017,Iaria.2005}. This fact implies that GX 9+1 is a sub-luminous Atoll source.

The next frontier in studying \gx~lies in exploring its polarimetric properties using the Imaging X-ray Polarimetry Explorer \citep[\ixpe;][]{Soffitta21,Weisskopf.2022,Weisskopf2023}, a joint NASA and ASI mission. \ixpe is equipped with three X-ray telescopes that utilize polarization-sensitive imaging gas-pixel detectors \citep{Costa.2001} operating in the 2--8 keV band. Several weakly magnetized X-ray binaries have been observed by \ixpe, providing valuable insights into their emission mechanism. These include the Z-class sources, such as Cyg X--2 \citep{Farinelli.etAl.2023}, XTE~J1701--462 \citep{Cocchi.etAl.2023}, GX 5--1 \citep{Fabiani.etAl.2023}, and Sco X-1 \citep{LaMonaca.etAl.2024}, as well as Atoll sources like GS 1826--238 \citep{Capitanio.etAl.2023}, GX 9+9 \citep{Ursini.etAl.2023}, 4U 1820--303 \citep{DiMarco.etAl.2023.4U,Anitra4U1820}, Ser~X--1 \citep{Ursini2024ser}, 4U 1624--49 \citep{Gnarini2024} and GX 3+1 \citep{Gnarini2024GX3p1}.
Polarimetric measurements have shown that Z-sources can reach polarization degrees of up to 4--5\%. This polarization is very variable and strongly related to the position of the source in the CCD. On the other hand, Atoll sources typically exhibit lower and less variable polarization levels ($\le$ 3\%). For most of the Atoll sources observed by \ixpe, polarization tends to increase with energy. For example, \cite{DiMarco.etAl.2023.4U} reported an unexpected spike in the polarization degree up to 9--10\% between 7 and 8 keV energy range in the ultra compact source, 4U 1820–303. 
Finally, as recently reported by \citet{Gnarini2024} for the Atoll source 4U~1624--49, the polarization signal is consistent with the Eastern-like scenario where Comptonization occurs within a boundary or spreading layer near the NS's surface, combined with the reflection of soft photons from the accretion disk. 

This paper presents a spectro-polarimetric study of \gx~using data from an observation campaign of various space missions (\nicer,~\nus,~\integral) performed simultaneously with \ixpe. The structure of the paper is as follows: in Sect. \ref{sec:Observations}, we describe the reduction of the data taken from \ixpe, \nus, \nicer and \integral observations; in Sect. \ref{Sec:data_and_results}, we present the data analysis and the results found. Finally, in Sect. \ref{sec:Disc_and_conc}, we discuss these results and the main conclusions.  
   \begin{table*}[ht]
   \centering
      \caption[]{Log of observations of \gx .}
         \label{log}       
\begin{tabular}{lccc}
            \hline
            \hline
            \noalign{\smallskip}
            Satellite  &  Obs ID & Start time (UTC) & Exp. Time (ks) \\
            \noalign{\smallskip}
            \hline
            \noalign{\smallskip}
            \ixpe &  03003801 & 2024-08-31~21:30:59 &   20.46  \\
            \nicer & 7700010101 & 2024-08-31~00:55:20 & 4.77  \\ 
            \nicer & 7700010102 & 2024-09-01~00:10:18 & 0.193  \\ 
            \nus & 31001009002 & 2024-08-31~18:41:11 & 10.65  \\
            INTEGRAL & rev.$\#$ 2817-2821 & 2024-08-28~01:30:41 &  188       \\
            \noalign{\smallskip}
            \hline
         \end{tabular}
   \end{table*}

\section{Observation and data reduction}\label{sec:Observations}
Table~\ref{log} reports the observation logs of all the instruments used in the simultaneous spectro-polarimetric campaign of \gx~. The source was observed simultaneously or quasi-simultaneously by \ixpe, \nicer, \nus, and we also found in the \integral data archive serendipitously simultaneous IBIS observations for a total of 188 ks. All the data were processed and analyzed using \textsc{heasoft} 6.33; \textsc{xspec}, version 12.14.1 \citep{Arnaud.1996} and the latest available calibration files. In the following, we describe in detail the data reduction of the instruments used in the data analysis.

\subsection{\ixpe data}
\ixpe observed \gx~on 2024 August 31, accumulating a total exposure time of 20.46 ks. We processed the cleaned level 2 event files using the standard \ixpe \textsc{ftools} procedure with the latest available calibration files and response matrices. We applied the weighted analysis method \citep{Baldini.etAl.2022,DiMarco.etAl.2022} for 
polarimetric data, utilizing the \texttt{stokes=Neff} parameter in \textsc{xselect}.
Source spectra and light curves were extracted from a circular region with a 120 arcsec radius, centered on, while the background was selected as an annular region spanning 180–240 arcseconds. We iteratively determined the source extraction radii from 30 to 180 arcseconds in 5-arcseconds steps in order to maximize the signal-to-noise ratio (S/N) across the entire IXPE energy range. This approach is similar to that used in previous analyses \citep[e.g.][]{Piconcelli.etAl.2004}. 

During \ixpe observation, the \gx~count rate was $>$ 10 cnt/s and varying around 30-40 cnt/s. Following \cite{DiMarco.etAl.2023}, background subtraction and rejection were not applied to the \ixpe spectra because it is unnecessary for sources with count rate $>$ 2 cnt/s.
For this fairly bright source, the gain fit is applied to properly account for a residual charging effect of the detectors not fully corrected by the pipeline (see Sect.~\ref{joint-spec}).
We analyzed the Stokes $I$, $Q$, and $U$ spectra independently for each detector unit (DU), employing a constant energy binning of 0.2 keV for the $Q$ and $U$ spectra, while the $I$ spectra were rebinned considering the optimal binning grouping by \cite{Kaastra.Bleeker.2016} with a minimum S/N of 3 for each bin using \texttt{ftgrouppha}.

\subsection{\nicer data}\label{nicerdata}
The Neutron Star Interior Composition Explorer (\nicer, \citet{Gendreau.etAl.2016}) performed two observations of the source around the \ixpe observations. The first observation started on 2024-08-31 at 00:55:20 UTC for 3.9 ks, while the second one started on 2024 September 1 at 00:10:18 for 193 s. Only a very short part of the observation overlaps with \ixpe for a total of only 350 s. %
The \nicer data were reduced using the \texttt{nicerl2} task to apply standard calibration and screenings, with CALDB version 20240206. Since \nicer observations were made during the orbital day, the \texttt{nicerl2} task was performed using also the \texttt{threshfilter=DAY} keyword\footnote{\url{https://heasarc.gsfc.nasa.gov/docs/nicer/analysis_threads/nicerl2/}}, otherwise the data would have been automatically screened by the \nicer software due to optical light leak problems. We extracted the source spectra with the \texttt{nicerl3-spect} command and the light curves using \texttt{nicerl3-lc}. The background was estimated with the SCORPEON\footnote{\url{https://heasarc.gsfc.nasa.gov/docs/nicer/analysis_threads/scorpeon-overview/}} model.  

We combined the \nicer spectra from the two observations and selected the good time intervals (GTIs) for the period overlapping with \ixpe and the entire \nus observation. A systematic error of 1\% was added to the 647 s obtained spectrum.

\subsection{\nus data}
\nus \citep{Harrison.etAl.2013} observed with its two X-ray telescopes on Focal Plane Modules A and B (FPMA and FPMB) over a net exposure time of 10.7 ks. Data were processed using the standard \texttt{nupipeline} task and the latest available calibration files (20240812). Since the source is bright (> 100 counts s$^{-1}$), the \texttt{statusexpr="(STATUS==b0000xxx00xxxx000)\&\&(SHIELD= =0)"} keyword was considered during the \texttt{nupipeline} task. Due to the presence of significant background across all energy bands, background subtraction was carried out for both detectors. The source extraction regions in the \nus images were defined by a circle with a radius of 160 arcsec, centered on \gx~, while background regions with a radius of 60 arcsec were selected from areas sufficiently far from the source. 

The source extraction radii were determined, as in the case of \ixpe, using an iterative method aimed at maximizing the S/N in the entire \nus energy range. The spectral data were then rebinned using the \texttt{ftgrouppha} task, following the optimal binning method outlined by \citet{Kaastra.Bleeker.2016}, with a minimum S/N of 3 per bin. FPMA and FPMB spectra were analyzed separately and not co-added for consistency across detectors. Data above 30 keV were excluded due to background dominance at higher energies.

We selected different \nus GTI to check the spectral characteristics: the GTI of the whole observations; the GTI so that the \nus observation lies within the \ixpe observation; the GTI divided based on the hard and soft colors (see \ref{Sec:timing}). 

\subsection{INTEGRAL data}  
\integral observed the source serendipitously for a total of 188 ks simultaneously with \ixpe. The \integral data were reduced using the latest release of the standard On-line Scientific Analysis (OSA, version 11.12), distributed by the INTEGRAL Science Data Centre \citep[ISDC,][]{Courvoisier03} through the multi-messenger online data analysis platform \citep[MMODA,][]{Neronov21}. The IBIS spectra were extracted in the 30--150 keV range using a response matrix with 256 standard channels. Only IBIS, the $\gamma$-ray energy detector \citep{Ubertini2003, Lebrun03}, was used for the data analysis to achieve a broader energy range in the spectra. However, IBIS did not detect the source with a 3$\sigma$ upper limit on the flux of $\sim 3\times10^{-11}\,\mathrm{erg\,cm^{-2}\,s^{-1}}$  in the 28--60 keV energy range. %

   \begin{figure}
   \centering
\includegraphics[width=9cm]{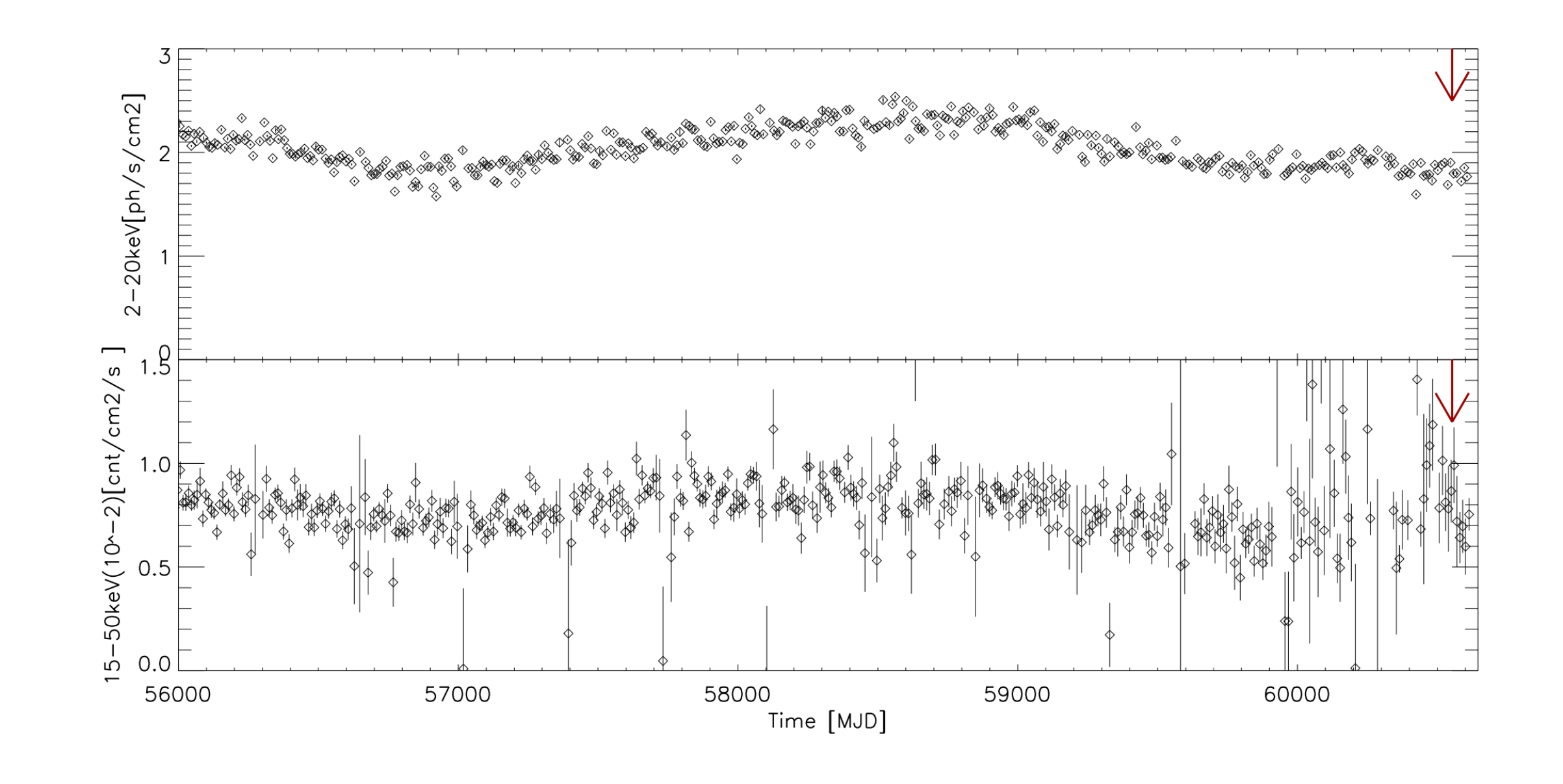}
      \caption{\gx~long term light curves at different energy ranges: 2--20 keV with MAXI (top) and 15--50 keV with BAT (bottom). The arrow indicates the times of the simultaneous \ixpe, \nus, \nicer, and \integral observation campaign.}
         \label{fig:bat_maxi}
   \end{figure}
   \begin{figure}
   \centering
\includegraphics[width=8cm]{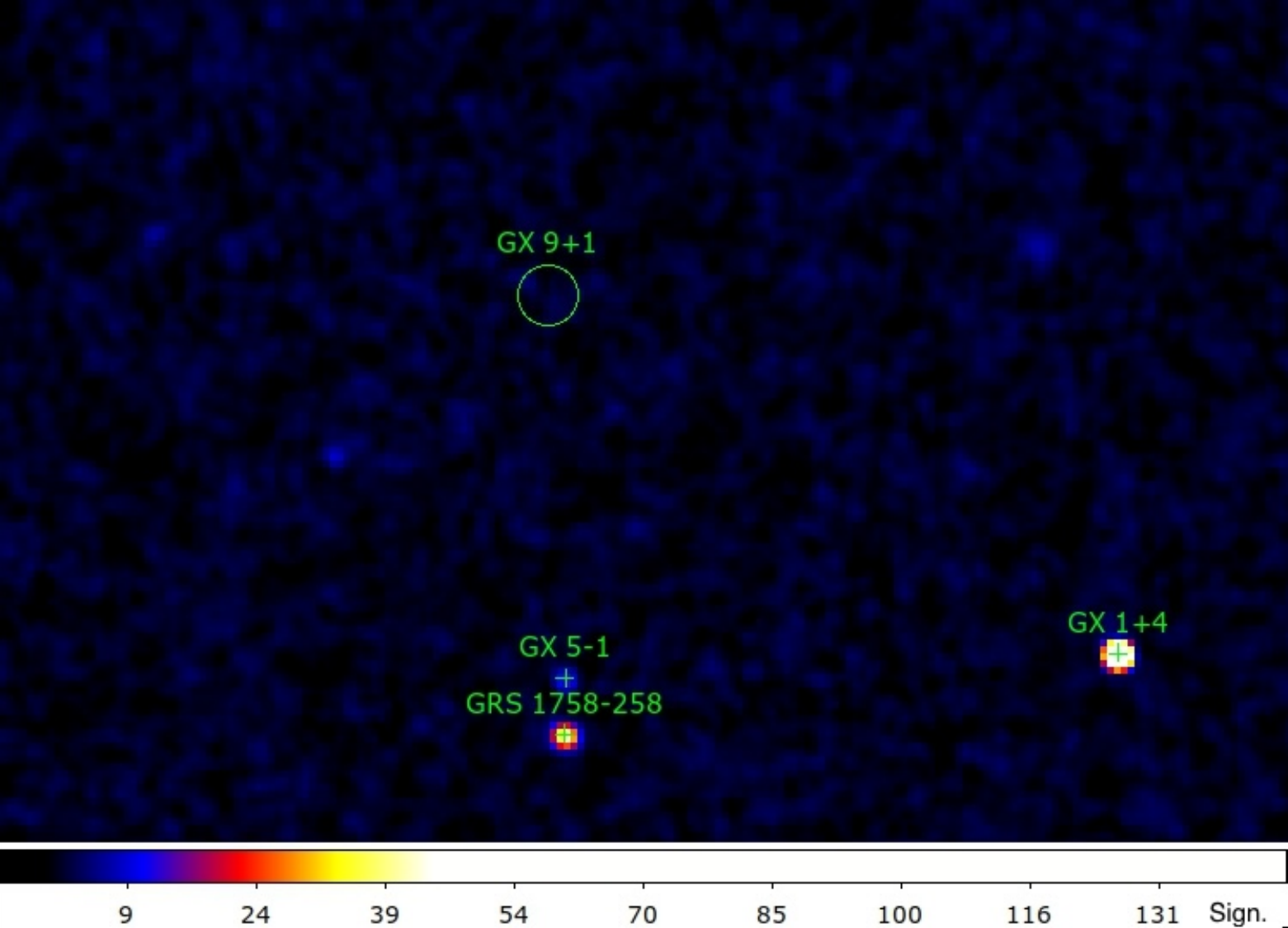}
      \caption{\integral/IBIS 28--60 keV mosaic image centered on \gx~of the data simultaneous with the \ixpe observation. The colobar indicates the significance.}
         \label{fig:integral}
   \end{figure}
\section{Data analysis and results}~\label{Sec:data_and_results}
\subsection{Timing behavior}~\label{Sec:timing}
Figure~\ref{fig:bat_maxi} shows the long-term \maxi and \bat light curves of \gx~. The pink arrows indicate the times of the simultaneous \ixpe, \nus, \nicer, and \integral observation campaign. As evident from the two light curves, the observation campaign was performed during a periodic minimum in the \gx~flux. Unfortunately, the \bat light curve is too noisy during the observation period to deduce the state of the source in the 15--50 keV energy range. The upper limit of 3.8 mCrab at 3$\sigma$ in the 28--60 keV \integral/IBIS mosaic image, Fig.~\ref{fig:integral}, indicates that the source's high-energy emission is concentrated below 30 keV. Above 30 keV, the source is very faint, confirming the absence of a hard tail, consistent with previously reported results \citep{Paizis2005}.
%
   \begin{figure}
   \centering
   \includegraphics[width=8cm]{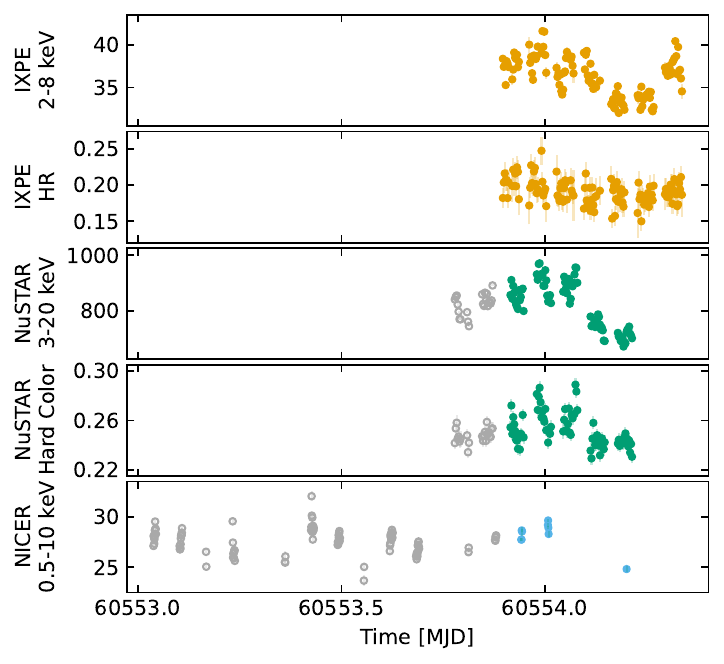}
      \caption{\ixpe, \nus (200 s time bins for both) and \nicer (50 s time bins) light curves of \gx (count s $^{-1}$). The second and fourth panels show the \ixpe and \nus hardness-ratios: 5--8 keV/3--5 keV and 10--20 keV/6--10 keV, respectively. Empty gray circles indicate points that are not simultaneous with the IXPE exposure.}
         \label{FigLC}
   \end{figure}
   \begin{figure*}
   \centering
   \includegraphics[width=16cm]{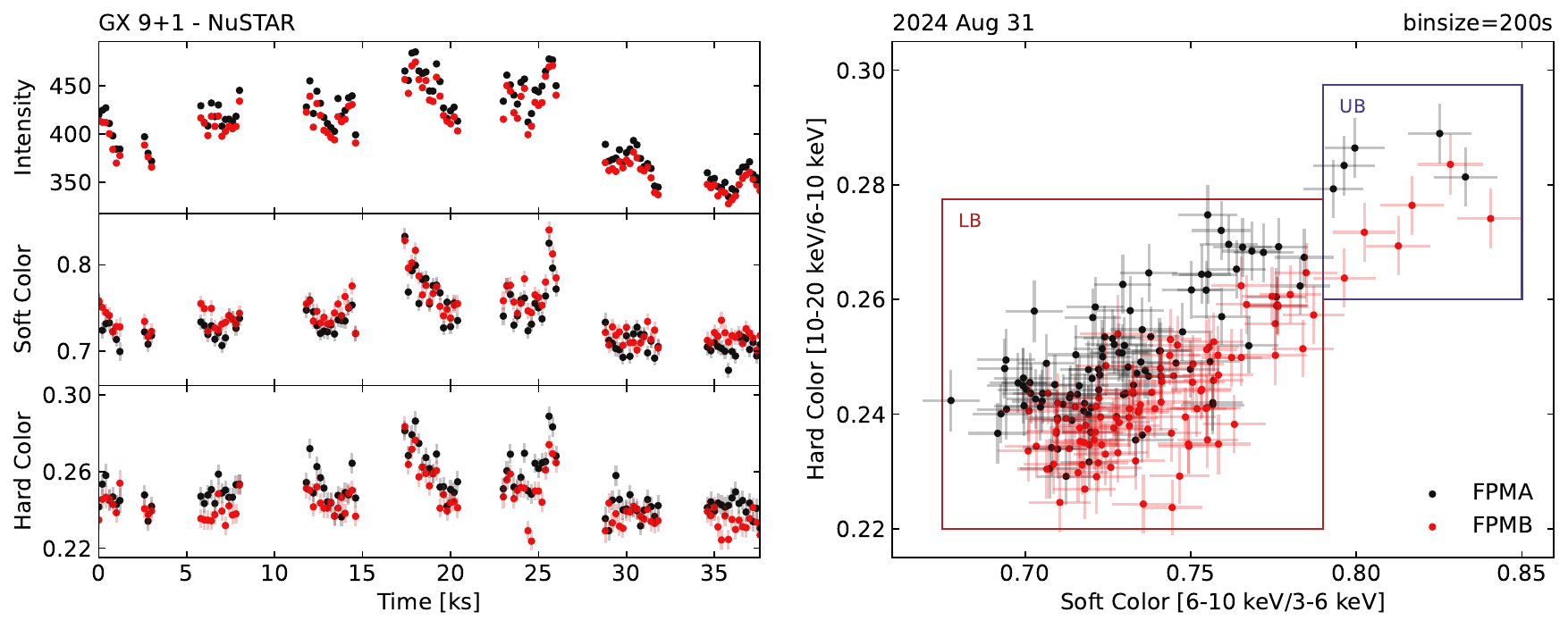}
   \hspace{-0.5cm}
   \caption{left panel: \nus~ 3--20 keV light curve, soft color 6--10 keV/3--6 keV and hard color 10--20 keV/6--10 keV of \gx. Right panel: \nus data color-color diagram of \gx. The Lower Box (LB) and Upper Box (UB) indicate the data grouped together in the same spectrum. See text for details.}
              \label{FigCCD}%
    \end{figure*}
%
Figure~\ref{FigLC} shows the light curves of \gx~ as observed by \ixpe, \nus, \nicer, along with the \ixpe hardness-ratio and the \nus hard color. The \ixpe hardness-ratio is defined as the 5--8 keV/3--5 keV flux ratio, while \nus hard color is calculated as the 10--20 keV/6--10 keV flux ratio. The time bins used are 200 s for \nus and \ixpe and 50 s for \nicer.
Figure~\ref{FigCCD} highlights the \nus light curves and hardness and the corresponding CCD. The soft color is calculated as the 6--10 keV/3--6 keV flux ratio. It is evident from Fig.~\ref{FigCCD} that the source during the \nus observation moves toward the banana branch. Based on the CCD value, we divided the points into two groups to check for spectral variability. The black "Upper Box" (indicated as UB, with a soft color > 0.79) include the harder points which are in the upper right part of the CCD. They are very few and correspond to the higher flux peaks in the light curves in Fig.~\ref{FigLC}. This confirms the well-known behavior of \gx~, where hardening corresponds to increase of the flux. The spectral parameters of the UB and those of the red "Lower Box" (indicated as LB, with a soft color <0.79) both correspond to a banana state and do not differ significantly enough to justify splitting the spectral analysis into two separate parts. Therefore, for the purposes of this study, we have chosen to combine all data into a single averaged spectrum to improve the S/N.

\subsection{Spectral analysis}\label{joint-spec}

We performed the joint \nicer+\ixpe+\nus broad band spectral analysis starting from the \nus spectrum alone, then adding \nicer and and finally incorporating the IXPE data at the end of the best-fit procedure.  
For the simultaneous fitting procedure, we used the entire \nus and \ixpe spectra and only the simultaneous part of the \nicer spectra with \nus (see Fig.\ref{FigLC}). 

For interstellar medium absorption, we adopt the most recent ISM abundances as described in \cite{Wilms.etAl.2000}. The $N_{\text{H}}$ was left free to vary during the fitting procedure.

We fit the \nus and \nicer spectra with a model composed of an absorbed disk black body (\texttt{diskbb}) plus a Comptonizing component. For the Comptonizing component we used a convolution model (\texttt{thcomp}) applied to the star/spreading-layer black body (\texttt{bbodyrad}). Before the fitting procedure, we extended the energy vector over which \texttt{thcomp} model is computed with the \textsc{xspec} command \texttt{"energies 0.01 1000.0 1000 log"}. 
However, taking into account the residuals of the fitting procedure, it is possible to detect a weak iron line around 6.2 keV. 
We tested a \texttt{Gaussian} component to model the presence of the iron line in our spectral fit. The Akaike Information Criterion (AIC) check indicates a significant improvement in the fit, with a $\Delta AIC$ of 12.54, supporting the necessity of this additional component. The AIC measures the information loss by using a specific model respect to another one \citep{Akaike1974AIC, Burnham.2004}\footnote{The model with the smallest AIC ($AIC = 2 \times n +\chi^2$) should be preferred, where $n$ is the number of free parameters of the fit. Generally, a difference of $\Delta AIC>10$ indicates a large and statistically significant improvement in the model's fit and the model with the lower AIC is considered to be substantially better than the other model. In the range $4 \leq\Delta AIC\leq 7$ the difference is not large enough to consider it a conclusive improvement. In this range, there is some evidence that one model is better, but the improvement is not drastic \citep{Burnham.2004}.}. Then, we also test a more complete and physical model, \texttt{RelxillNS}, which may take into account the presence of a small amount of reflection component.
The AIC test in this case shows that statistically there is no preference over the simple gaussian model ($\Delta AIC= 4.22$). However, we chose to use the \texttt{RelxillNS} model despite its low significance and flux contribution, as it offers more detailed insights into the physical parameters.
The \texttt{RelxillNS} model package reproduces the relativistic reflection from the innermost regions of an accretion disk \citep{Garcia.etAl.2014, Dauser.etAl.2014}. Specifically, \texttt{RelxillNS} assumes a single-temperature blackbody spectrum that illuminates the surface of the accretion disk at 45\degr, which could physically originate from the emission of the spreading layer (or NS surface) \citep{Garcia.etAl.2022}. We fixed the reflection fraction parameter in \texttt{RelxillNS} to a negative value in order to obtain only reflected component as indicated in \citet{Garcia.etAl.2016}. We also linked the temperature of the seed photons to the temperature in the bbodyrad component. We fixed the number density at $\log(n_{\rm e})$ = 18, as reported in \citet{Garcia.etAl.2016}, the emissivity index to the best fit found ($q_{\rm em}$=3), the dimensionless spin $a$ of the NS to 0.2 that is a typical spin frequency \citep[see e.g.,][]{Braje.etAl.2000}. 

\nicer data revealed residuals below 2.5 keV in the energy spectra, due to features not corrected in the \nicer ancillary response file (ARF) \citep{Miller.etAl.2013,Strohmayer.etAl.2018}. To account for this, we need to include three absorption edges to the \nicer data (\texttt{edge} in \textsc{xspec}) at 1.5, 1.8 and 2.4 keV, respectively. It should be noted that numerous edges detected in previous observations have also been directly attributed to the source, likely due to the presence of ionized material around the system \citep{Thomas2023, Iaria.2005}.

\begin{figure}
   \centering
   \includegraphics[width=9cm]{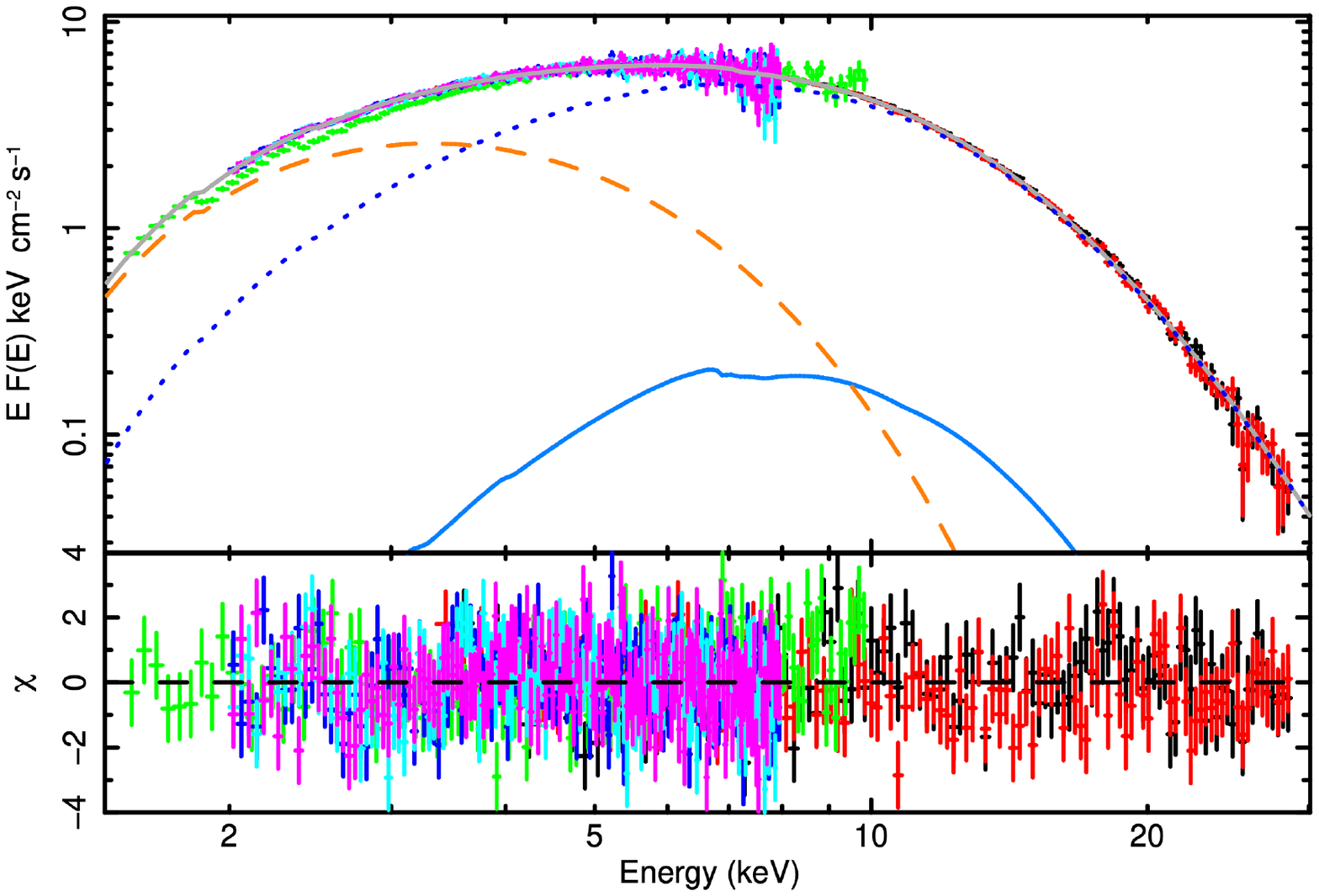}
   \caption{\nus~(red and black points), \nicer~(light green points) and \ixpe-I (light blue, magenta and violet points for each DU, respectively) unfolded spectrum of \gx. Dotted line: Comptonization component; dashed line: disk component; solid line: reflection.}
              \label{fig:Spe_NUSTAR-IXPEI-NI}
    \end{figure}

The obtained joint \nicer+\ixpe+\nus energy spectrum is reported in Fig.~\ref{fig:Spe_NUSTAR-IXPEI-NI}. The \textsc{xspec} syntax of the used model is: \texttt{constant*edge*edge*edge*TBabs(diskbb+relxillNS+\\thcomp*bbodyrad)}. Table~\ref{table:fit_nustar} reports the spectral parameter values obtained from the joint \nicer+\ixpe+\nus spectral fit. Table~\ref{table:const} reports the cross-calibration constants between the different instruments, determined by setting the \nus FPMA constant to 1, the gain shift of IXPE and the NICER edges values. Both tables values are reported with 90\% confidence level (CL), such as $1.6\sigma$. The normalizations of the \texttt{diskbb} and \texttt{bbodyrad} components are provided assuming a source distance of 10 kpc. The fit results in reasonable values of the spectral parameters for a LMXB hosting a NS as compact object, and indicates that \gx~ is in a banana state observed at relatively low inclination angle. The low Fe abundance indicates a late-type companion star. 

The obtained $N_{\text{H}}$ value could be probably overestimated even if compatible with previous reported study \citep{vanderBerg2017,Thomas2023}. In fact, as reported by \cite{Iaria.2005}, the BeppoSAX spectra of \gx\/ show absorption edges due to the presence of material in front of the source.
For this reason, we attempted to account for variations in iron and oxygen abundances by applying the Tübingen-Boulder ISM absorption model (\texttt{TBfeo} in \textsc{xspec}) to the total emission. Unfortunately, we found this model difficult to constrain to physical acceptable oxygen and iron abundances, likely due to both the non-standard element abundances and the short duration of the simultaneous \nicer spectrum. In any case, the other spectral parameters are not affected by the choice of different absorption models.

 Additionally, Table~\ref{table:const} provides the flux values of the source in different energy bands, as well as the ratios of the fluxes of the different model components relative to the \ixpe bands 2--8 keV and 2--3 keV, respectively.
The source luminosity is calculated as $L/L_{\rm Edd}=1.5$\%, assuming a mass of 1.4 $M_{\odot}$ and a source distance of 4 kpc ~\citep{vanderBerg2017,Iaria.2005}. This leads to an estimated accretion rate of $\dot{m}\sim 4.5\times 10^{-10}$ $M_{\odot}$/yr, assuming an accretion efficiency of 0.1.

\begin{table}[ht]
\centering
\caption{Spectral fitting (Part 1): model fitting parameters \nicer, \nus, and \ixpe. The \textsc{xspec} syntax of the used model is: \texttt{constant* edge*edge*edge*TBabs(diskbb+relxillNS+thcomp*bbodyrad)}. Square brackets indicate fixed values. Values are reported at 90\% CL. See also Table 3.}
\begin{tabular}{l r l}
\hline
\hline
Model & Parameter & Value \\
\hline
\multirow{1}{*}{\texttt{TBabs}} & $N_{\text{H}}$ ($10^{22}$ cm$^{-2}$) &  $2.72^{+0.06}_{-0.03}$ \\
\hline
\multirow{2}{*}{\texttt{diskbb}} & $kT_{\text{in}}$ (keV) & $1.08^{+0.03}_{-0.02}$ \\ 
                                & $R_{\text{in}} \sqrt{\cos i}$ (km) & $19.0^{+0.8}_{-0.7}$\\ 
\hline
\multirow{2}{*}{\texttt{bbodyrad}} & $kT_{\text{bb}}$ (keV) & $1.68^{+0.02}_{-0.02}$ \\
                         & $R_{\text{bb}}$ (km) & $11.4^{+0.5}_{-0.5}$ \\
\hline
\multirow{3}{*}{\texttt{thcomp}} & $\tau$ &    $4.8^{+1.1}_{-1.2}$\\
                         & $kT_{\text{e}}$ (keV) & $3.4^{+0.5}_{-0.3}$ \\
                        & $cov_{\text{frac}}$ & $0.5^{+0.1}_{-0.1}$ \\
\hline
\multirow{9}{*}{\texttt{relxillNS}} & $q_{\text{em}}$ & [3]\\
 & $a$ & [0.2] \\
 & Incl (deg) & $34^{+12}_{-12}$ \\
 & $R_{\text{in}}$ (RISCO units) & $<1.2$ \\
 & $kT_{\text{bb}}$ (keV) & [=$kT_{\text{bb}}, \text{bbodyrad}$] \\
 & $\log(\xi/\text{erg cm s}^{-1})$ & $1.5^{+0.6}_{-0.4}$\\
 & $A_{\text{Fe}}$ & $<1.1$  \\
 & $\log n_{\text{e}}$ & [18] \\
 & $N_{\text{r}}$ ($10^{-3}$) & $1.9^{+1.2}_{-1.2}$ \\
 \hline
 &$\chi^2$/d.o.f. & 887.41/769$\simeq$1.13 \\
\hline
\label{table:fit_nustar}
\end{tabular}
\end{table}
%
\begin{table}[ht]
\centering
\caption{Spectral fitting (Part 2): cross-calibration constants, \ixpe gain shift, NICER absorption edges (Energy, $E$,  and absorption depth at the threshold, $d$), unabsorbed photon fluxes, and flux ratios. The flux values are given with an error within 1$\%$. Values are reported at 90\% CL.}
\begin{tabular}{l l}

\hline
\hline
Parameter & Value \\
\hline
$C_{\text{FPMB-FPMA}}$ & $0.981^{+0.001}_{-0.001}$ \\
$C_{\text{XTI-FPMA}}$ & $0.940^{+0.004}_{-0.004}$ \\
$C_{\text{DU1-FPMA}}$ & $0.808^{+0.003}_{-0.003}$ \\
$C_{\text{DU2-FPMA}}$ & $0.809^{+0.003}_{-0.003}$ \\
$C_{\text{DU3-FPMA}}$ & $0.790^{+0.003}_{-0.003}$ \\
\hline
$\alpha_{\text{DU1}}$ & $1.017^{+0.004}_{-0.004}$ \\
$\beta_{\text{DU1}}$ (eV) & $-42^{+15}_{-16}$ \\
$\alpha_{\text{DU2}}$ & $1.008^{+0.004}_{-0.005}$ \\
$\beta_{\text{DU2}}$ (eV) & $-24^{+15}_{-18}$ \\
$\alpha_{\text{DU3}}$ & $1.016^{+0.004}_{-0.004}$ \\
$\beta_{\text{DU3}}$ (eV) & {$-24^{+16}_{-15}$} \\
\hline
$E_1$ (keV) & 1.52$^{+0.03}_{-0.05}$ \\
$d_1$ & 0.16$^{+0.02}_{-0.02}$ \\
$E_2$ (keV) & 1.85$^{+0.02}_{-0.02}$\\
$d_2$ & 0.17$^{+0.02}_{-0.02}$ \\
$E_3$ (keV) & 2.38$^{+0.03}_{-0.03}$\\
$d_3$ & 0.11$^{+0.01}_{-0.02}$ \\
\hline
\multicolumn{2}{c}{Fluxes} \\
                  $F_{\text {E range (keV)}}$ & (erg s$^{-1}$ cm$^{-2}$) \\
\hline
\hline
$F_{\text{2--8}}$ & $ 1.27\times 10^{-8}$ \\  
$F_{\text{2--3}}$  & $ 3.10\times 10^{-9}$ \\ 
$F_{\text{3--8}}$  & $ 9.60\times 10^{-9}$ \\ 
$F_{\text{3--30}}$  & $ 1.38\times 10^{-8}$ \\ 
$F_{\text{bol}}$  & $ 2.36\times 10^{-8}$ \\ 
\hline
\hline
$F_{\texttt{diskbb}}/F_{\text{2--8}}$ & 41.34 \% \\
$F_{\texttt{(thcomp*bbodyrad)}}/F_{\text{2--8}}$ & 57.09 \% \\
$F_{\texttt{relxillNS}}/F_{\text{2--8}}$ & 1.61 \% \\
\hline
$F_{\texttt{diskbb}}/F_{\text{2--3}}$ & 69.78 \% \\
\hline
\end{tabular}
\label{table:const}
\end{table}

\subsection{Spectro-polarimetric analysis}~\label{sec:Spec_pol}
\begin{figure}
    \centering
    \includegraphics[width=\linewidth]{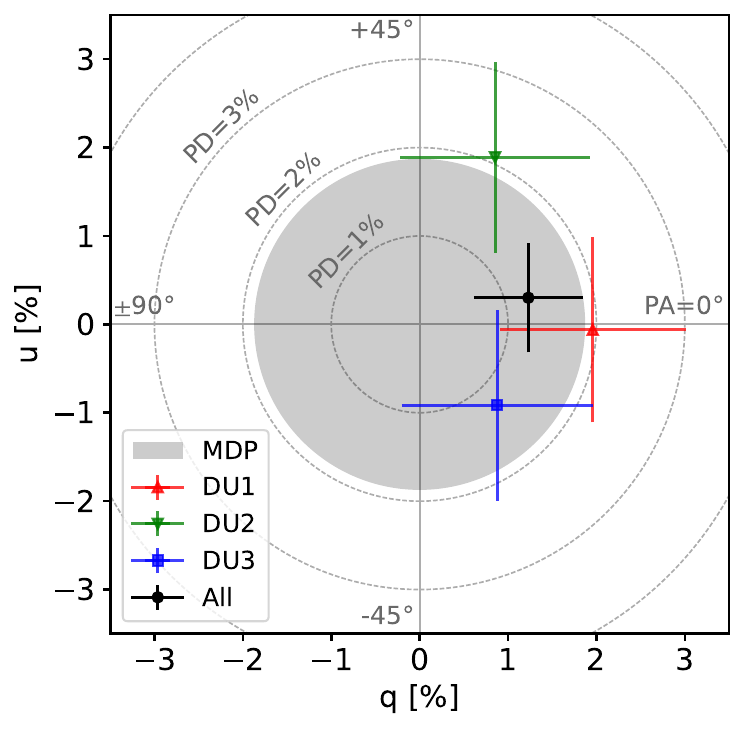}
    \caption{2--8 keV normalized Stokes $q$ ($Q/I$) and $u$ ($U/I$) parameters obtained for the three DUs with the \texttt{PCUBE} algorithm of \textsc{ixpeobssim} \citep{Baldini.etAl.2022}. The black cross represents the combination of the three DUs, while the gray-filled circle corresponds to the combined 99\% MDP.}
    \label{fig:PCUBE.Stokes}
\end{figure}

Firstly, we performed the polarimetric unweighted analysis of the entire \ixpe observation using \textsc{ixpeobssim} (v.31.0.1; \citealt{Baldini.etAl.2022}), with the latest available calibration file (v.13.20240701). Figure  \ref{fig:PCUBE.Stokes} shows for each DU the normalized Stokes parameters measured by \ixpe and obtained using the \texttt{PCUBE} task of \textsc{ixpeobssim} with the minimum detectable polarization (MDP) at the 99\% level. The black cross representing the combination of the three DU signals is below the 99\% MDP. Therefore, we can only provide an upper limit for the PD in the 2--8 keV range, and 3.0$\pm$0.9 \% at 3.1 $\sigma$ in the 2--3 keV.
%
%
\begin{table}[h]
    \caption{Polarization degree and angle (PD, PA) values at different energy bands (band) and confidence levels (CL) for 1 parameters with \textsc{xspec} considering the model: \texttt{polconst*constant*edge*edge*edge*Tbabs(diskbb+relxillNS\\+thcomp*bbodyrad).}}
    \label{tab:polarization}
    \centering
     \begin{tabular}{crlc}
        \hline 
        \hline
           band (keV) & PD ($\%$)    &   PA ($^\circ$)   &    PD ($\%$) \\       
          & \multicolumn{2}{c}{90$\%$ CL (1.6$\sigma$)} & 99.7$\%$ CL (3$\sigma$) \\
         \hline
        \vspace{0.1cm}
        2--3  & $2.5_{-1.5}^{+1.5}$ & $+2\pm18$ & < 5.2 \\
        \vspace{0.1cm}
        2--4  & $0.96_{-0.91}^{+0.91}$ & $-4\pm37$   &  < 2.6 \\
        \vspace{0.1cm}
        4--8  &  <1.5 & & < 2.5 \\
        2--8  & <1.4  & & < 1.9 \\
        \hline
\end{tabular}
\end{table}
 \begin{figure*}
   \centering
   \includegraphics[width=6cm]{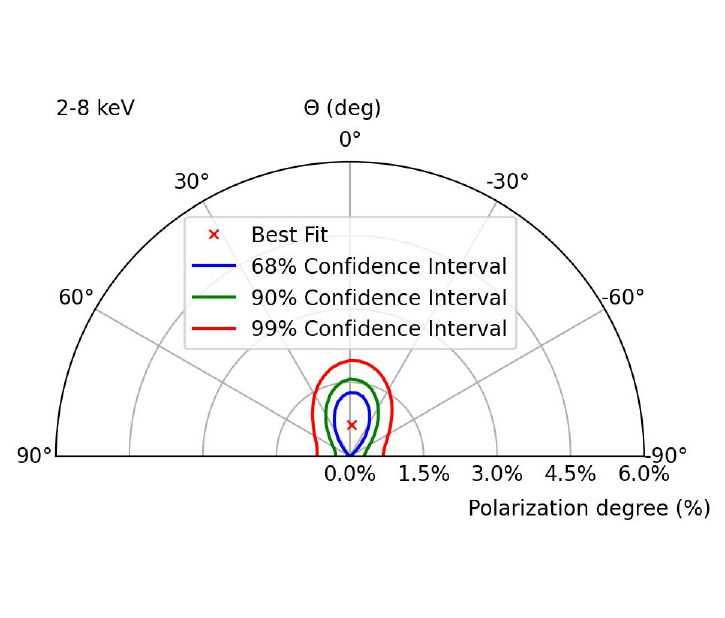}
   \includegraphics[width=6cm]{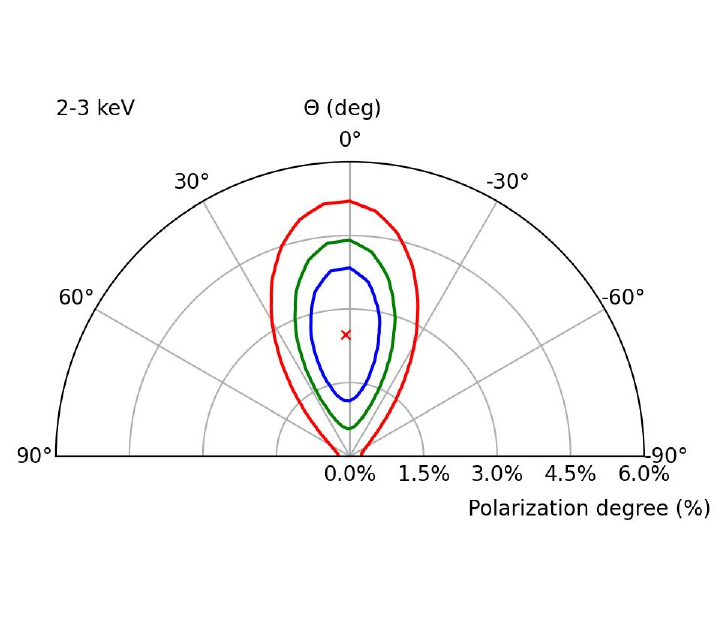}
 \includegraphics[width=6cm]{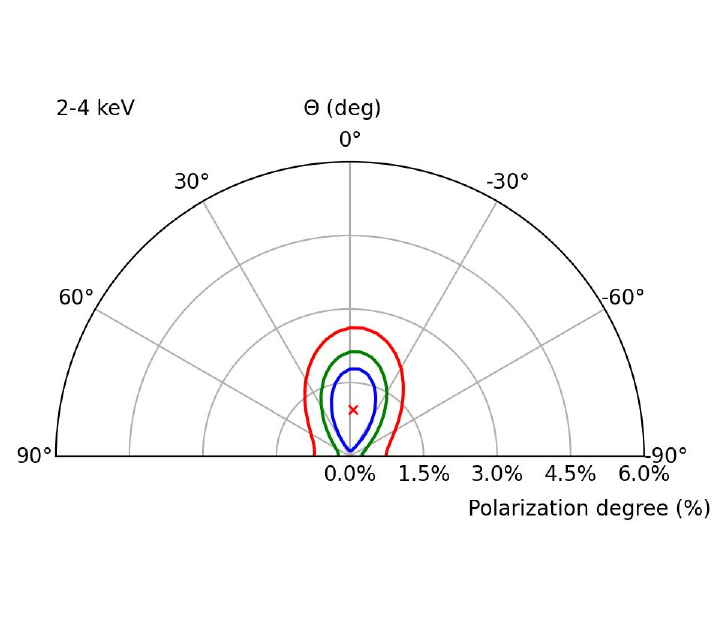}
 
 \includegraphics[width=6cm]{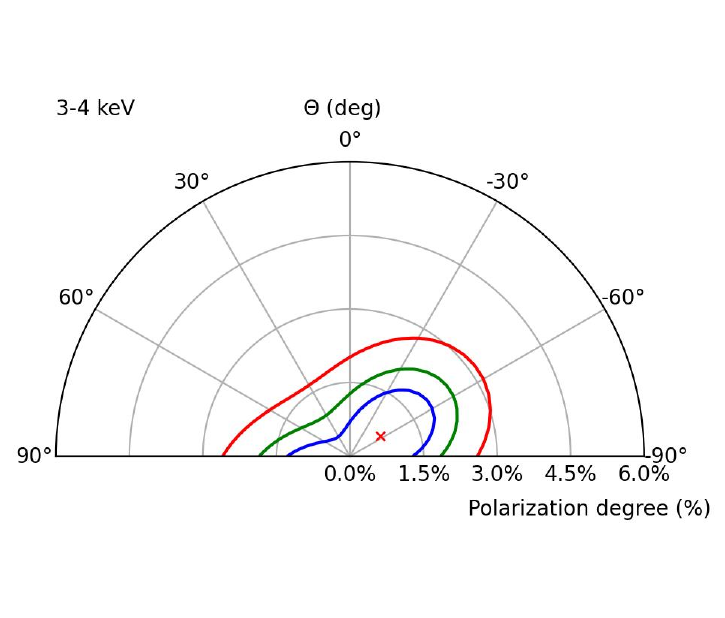} 
 \includegraphics[width=6cm]{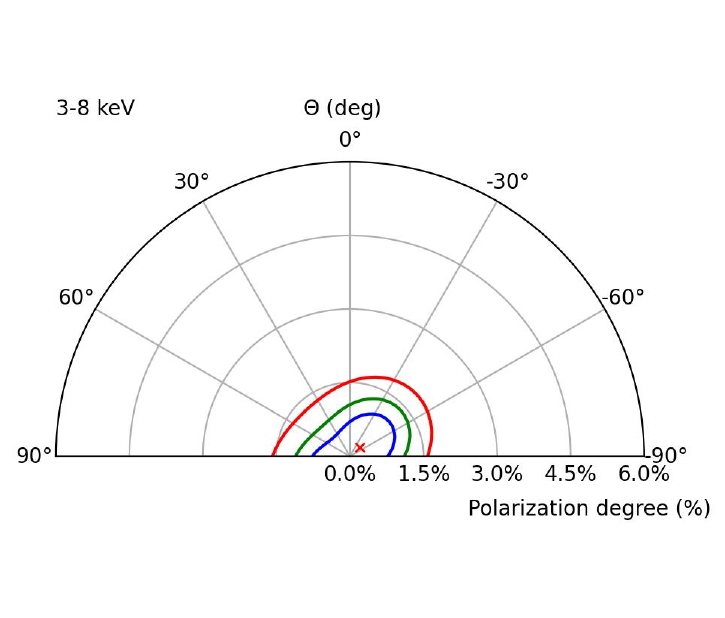}
 \includegraphics[width=6cm]{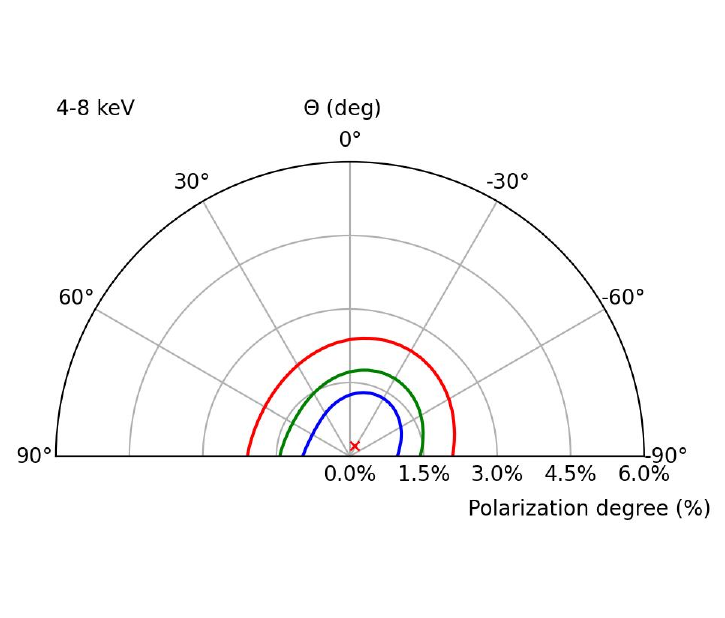}
   \caption{Contour plots of the PD and PA at the 68.3\% (1$\sigma$), 90\% ($1.6\sigma$), and 99\% confidence levels ($2.6\sigma$) obtained with \textsc{xspec} at different energy ranges.}
              \label{Fig:polplot}
    \end{figure*}

The spectro-polarimetric analysis was performed by adding the \texttt{polconst} convolution model to the total model and applying it to the entire set of \ixpe spectra ($I$, $Q$ and $U$), the \nicer spectrum and the two \nus spectra. The \textsc{xspec} syntax of the model applied is: \texttt{polconst*constant*edge*edge*edge*Tbabs(diskbb+\\relxillNS+thcomp*bbodyrad)}. 
In order to derive the polarization parameters (PD and PA of the \texttt{polconst} model), the other spectral parameters of the total model were fixed to those found from spectral fitting (see Table~\ref{table:fit_nustar}).
Table~\ref{tab:polarization} reports the results of the \textsc{xspec} analysis of PD and PA at different CL computed with the error command for one parameter of interest. The upper limit is reported at the 99.7\% CL (3$\sigma$).
Figure~\ref{Fig:polplot} reports the contours of PD and PA obtained with \textsc{xspec} of the \ixpe observation in the 2--3, 2--4, 3--4, 3--8, 4--8 and 2--8~keV energy bands. 
As Figure~\ref{Fig:polplot} shows, the PD between 2--8 keV is compatible with null polarization above 68.3\% CL (1$\sigma$), and consequently the PA is unconstrained. The same result is obtained by studying the PD and PA at different energy intervals with the exception of the 2--3 keV energy interval. In this case, we found a detection at the 95.5\% CL (2$\sigma$) in the energy range 2--3 keV and a hint of detection at the 68.3\% CL (1$\sigma$) in the 2--4 keV. Consistent results are also obtained using the model-independent \textsc{ixpeobssim} task (as stated the beginning of this section).

On the basis of these results, we also applied \texttt{polconst} to each individual spectral component. The \textsc{xspec} syntax of the applied model is: \texttt{constant*edge*edge*edge*TBabs(polconst*diskbb+\\
polconst*relxillNS+polconst*(thcomp*bbodyrad))}. We fixed the PA of the reflection component to the best fit of the Comptonized component, $90^\circ$, and allowed the PD values to vary. We obtained an upper limit for the Comptonized component of PD$<$5.1\% at 99.7\% CL, while the reflection component PD is not constrained $<$ 100\% at 99.7\% CL. Instead, the PD of the disk component converges to a PD value (PD=$2.1^{+2.4}_{-1.8}$\%) that is consistent with the one obtained in the 2--3 keV energy range when applying \texttt{polconst} to the entire model as Table~\ref{tab:polarization} reports. It is noteworthy that the best fit of the PA (PA=$-1.7^{+36}_{-37}$$^\circ$) is naturally orthogonal to the fixed PA of Compton and the reflection components, though uncertainties are large. We also attempted to fix the PD of the Comptonization component to the best-fit value of 0.39\%, which is a plausible value for low inclination angles as reported by \citet{Gnarini.etAl.2022}, while the PD of the reflection component was fixed at 10\% \citep{Lapidus.Sunyaev.1985}. Moreover, we assumed, as in the previous case, that the PA of the reflection and Comptonized components are aligned in the same direction, fixing the angle to $90^\circ$. The fitting procedure again returned a disk PD value consistent with the value obtained in the 2--3 keV energy range when applying \texttt{polconst} to the entire model and equal to PD--$\text{disk}$ = $4.6^{+1.3}_{-1.3}$\%) with a PA = $-0.74^{+8.4}_{-8.4}$ degree (90\% CL).

\section{Discussion and conclusions}~\label{sec:Disc_and_conc}

We report about the \ixpe observation of \gx~ with simultaneous \nus and \nicer observations, with serendipitous simultaneous \integral observations. 
This new observation campaign on this source offers new insights into the polarimetric and spectroscopic properties of this NS-LMXB system. \gx~ presents a unique comparison with other similar objects, especially given the differences in polarization behavior observed across the Atoll and Z-class sources. Previous studies of Atoll sources, such as GX 9+9 and GS 1826--238, have highlighted low values in polarization degrees, with GX 9+9 showing a PD of $\sim$ 2\%, linked to a combination of boundary layer (BL) and reflection effects \citep{Ursini.etAl.2023}, GS 1826--238 having only a strict upper limit less than 1.3\% in 2--8 keV. For \gx~, our results demonstrate that the polarization degree is relatively modest and less than $\simeq$1.9\% in 2--8 keV, as in the case of GS 1826--238, Ser X--1, GX 3+1 and in line with expectations for systems with low inclination angles.

The broad-band spectra of \gx~ can be described by a soft thermal component and a harder Comptonized emission, with characteristic spectral parameters of an Atoll source in the soft state. Moreover, also faint reflection features are present in the spectrum, their contribution to the total flux is very low ($\simeq$2\%) and even if the reflection component is probably highly polarized ~\citep{Lapidus.Sunyaev.1985} it does not contribute significantly to the total PD. The spectrum is fairly dominated by Compton emission at about 57\% of the total flux, while $\sim$ 41\% of the flux is due to the disk emission. 

However, the 2-8 keV PD of < 1.9\% is consistent with previous findings for other Atoll sources, where the polarization degree often does not exceed 2\%, except in cases where a significant Comptonized component, along with the reflection component, dominates the emission \citep[see for example the steep increase in PD between 7--8 keV in 4U 1820--30;][]{DiMarco.etAl.2023.4U} or the source is seen at a high inclination angle \citep[see for example the dipping source 4U 1624--49;][]{Saade.etAl.2024,Gnarini2024}.

We can claim a marginal detection between 2 and 3 keV energy range (at the 90\% CL; see Fig.~\ref{Fig:polplot} and Table~\ref{tab:polarization}). It is noticeable that in that lower energy range (2--3 keV) the disk emission represents 70\% of the total emission, while the remaining 30\% is mostly due to the Comptonization component. Our results indicate that most of the PD detected between 2--3 keV is due to disk emission. In fact, applying the constant polarization model to each component of the model (see Sect.~\ref{sec:Spec_pol}) in order to study their polarization, the PD of the disk component converges to a value compatible with the marginal detection found between 2--3 keV and the best fit of PA naturally converges to 90 degrees with respect to the Comptonized and reflection PA. The PD value found is quite high to be due to a standard accretion disk modeled as a semi-infinite plane-parallel atmosphere~\citep{Chandrasekhar.1960}, however, the 90\% uncertainty is very large. 

Moreover, as reported by \citet{Gnarini.etAl.2022}, simulations considering the shell geometry of the corona show an increase in PD below 3 keV. Even if this increase seems evident only for inclinations fairly higher than those found in our analysis (>$40^\circ$), it appears consistent within the errors with our findings.
The inclination angle of \gx~is still a subject of debate. However, our results estimate that the inclination of the source is quite low, $23\degr<i<46\degr$ (see Table \ref{table:fit_nustar}). This probably affects the overall polarization degree in agreement with the upper limit found on PD. As observed in sources such as Ser X-1 and GX 9+9, the disk inclination and the geometry of the Comptonizing region are key factors in determining the polarization properties. The detection of polarization at the 90\% CL implies that further longer observations of \gx~with \ixpe, possibly in conjunction with other observatories like \nicer or \nus, could easily detect a polarization signal, help clarify its nature, and refine the models used to interpret these data. The insights from polarimetric observations will also be crucial for understanding the connection between spectral states and the geometric configurations of these systems.

Moreover, as Table~\ref{table:fit_nustar} shows, the covering factor of the Comptonized component is quite low ($\sim 0.5$). This can be explained within the scenario outlined by the spectro-polarimetric analysis. In fact, the low inclination obtained from the spectral fit, along with the low polarization and low covering factor, can be well explained if we consider the spreading layer as the site of Comptonization of the NS seed photons. Observing the source at low inclination allows us to directly observe the pole of the NS.

\begin{acknowledgements}
This work reports observations obtained with the Imaging X-ray Polarimetry Explorer (IXPE), a joint US (NASA) and Italian (ASI) mission, led by Marshall Space Flight Center (MSFC). The research uses data products provided by the IXPE Science Operations Center (MSFC), using algorithms developed by the IXPE Collaboration, and distributed by the High-Energy Astrophysics Science Archive Research Center (HEASARC). AT, FC, and SF   acknowledge financial support by the Istituto Nazionale di Astrofisica (INAF) grant 1.05.23.05.06: ``Spin and Geometry in accreting X-ray binaries: The first multi frequency spectro-polarimetric campaign''. AG, FC, SB, SF, GM, PF and FU acknowledge financial support by the Italian Space Agency (Agenzia Spaziale Italiana, ASI) through the contract ASI-INAF-2022-19-HH.0. SF, PS have been also supported by the project PRIN 2022 - 2022LWPEXW - ``An X-ray view of compact objects in polarized light'', CUP C53D23001180006. MN is a Fonds de Recherche du Quebec – Nature et Technologies (FRQNT) postdoctoral fellow. JP thanks the Academy of Finland grant 333112 for
support.
This research used data products and software provided by the \ixpe, the \nicer and \nus teams and distributed with additional software tools by the High-Energy Astrophysics Science Archive Research Center (HEASARC), at NASA Goddard Space Flight Center.
\end{acknowledgements}
%
%

\bibliographystyle{aa} 

\end{document}